# Revealing Giant Planet Interiors Beneath the Cloudy Veil

**Tristan Guillot (Université Côte d'Azur, OCA, Lagrange CNRS, 06304 Nice, France, tristan.guillot@oca.eu)**
**Leigh N. Fletcher (School of Physics and Astronomy, University of Leicester, University Road, Leicester, LE1 7RH, UK, leigh.fletcher@le.ac.uk)**

**Observations from the Juno and Cassini missions provide essential constraints on the internal structures and compositions of Jupiter and Saturn, resulting in profound revisions of our understanding of the interior and atmospheres of Gas Giant planets. The next step to understand planetary origins in our Solar System requires a mission to their Ice Giant siblings, Uranus and Neptune.**

## Juno and Cassini unveil the depths of Jupiter and Saturn

Legend has it that Zeus (Jupiter) shrouded himself beneath a cloudy veil to hide his mischiefs, and that his wife Hera (Juno) was able to lift these clouds to confront him. The Juno mission, in orbit around Jupiter since 2016, indeed was able to lift some of the veil surrounding that planet. With a slightly different approach, the Cassini mission did the same at Saturn. But the task is proving more challenging than initially anticipated: Jupiter and Saturn are not eager to see all of their inner secrets so easily discovered.

Thanks to its special polar orbit, Juno has been able to measure Jupiter's gravity field with unprecedented accuracy, more than 2 orders of magnitude better than what had been achieved previously (1,2), a milestone for research on interior models of this archetypal giant planet. This led to the resolution of a long-standing mystery: would the strong banded pattern of zonal winds in their atmosphere (with speeds exceeding 100m/s) be limited to a shallow atmospheric layer, or would they penetrate deep into the interior? Juno has revealed that Jupiter's zonal winds extend down to about 3000 km below the visible clouds and that differential rotation becomes small at greater depths (3,4). Similar measurements from the Grand Finale orbits of Cassini (the last part of Cassini's mission before its final plunge into the giant planet in 2017) revealed that, as anticipated from the Juno results, Saturn's deep winds must extend deeper, to about 9000 km (5,6). This shows that atmospheric zonal winds extend relatively deep, until hydrogen becomes conductive enough to be dragged by the interior magnetic field into uniform rotation. This new insight helps us to understand the interiors, magnetic fields, and atmospheric dynamics both on our Gas Giants, and also on giant exoplanets.

The accurate gravity field determination also led to the realization that Jupiter's envelope is inhomogeneous (7,8): It contains more heavy elements somewhere within its metallic interior, about halfway inside the planet, than in its outer envelope. The presence of this so-called diluted core tells us that Jupiter, despite being internally hot and therefore mostly convective, is unable to fully mix its heavy elements.

A few years before, Cassini's discovery of free oscillations of Saturn through perturbations of its rings (9) led to a similar surprise: in order to explain the complex splitting of the modes, interactions between different types of oscillations is needed, indicating the existence of both the usual convective region where p-modes (for which the restoring force is pressure) propagate and a deep stable region where g-modes (with gravity as the restoring force) are present (10). In that case as well, the hypothesis of a fully convective interior led way to a more complex structure.

Thus, the interiors of these planets are not as simple as previously thought. They are only partially mixed. For Jupiter, the reason certainly lies in the initial conditions: if the planet formed with a large diluted core, only partial mixing would have occurred, with large composition gradients remaining present throughout the planet's lifetime (11). However, the causes of such a large diluted core are not clear, and the possibility that the planet witnessed a giant impact early in its history is, at present, the most likely explanation (12). In Saturn, the situation is different: the presence of a large stable region may be the result of an extensive helium phase separation leading to the formation of an almost pure-helium core (13). A detailed characterization of the deep interiors will require seismology. But progress will come also from a better coupling of interior, dynamo and atmospheric models constrained by the Juno and Cassini measurements.

## Ever-Changing Atmospheres

Giant planet atmospheres provide a lens through which we can catch a glimpse of the hidden depths. Peering through clearings in the clouds, visible and infrared observations reveal a complex story involving composition changes driven by vertical and horizontal motions (14). Conventional theoretical models predict the formation of clouds above a well-defined condensation level, with uniform mixing below that point. Radiometric observations, which penetrate through the clouds and can probe very deep compared to visible or infrared observations, provide a strikingly different picture. In fact, ground-based observations from the Very Large Array 30 years ago showed that ammonia is depleted across most of Jupiter except near its equatorial zone (15). Juno's MicroWave Radiometer (MWR) further demonstrated that Jupiter's ammonia has a variable abundance as a function of depth and latitude down to at least 200km below the cloud tops, far beneath the expected cloud base (1, 16). Juno's key contributions have been revealing the great depth of the ammonia depletion and the fact that it affects most of the planet.

With hindsight, the strong spatial variability in the sub-cloud layers is not altogether unexpected – the spatial distribution of disequilibrium species (like phosphine, arsine, germane, and para-hydrogen) and lightning (indicating the vigor of moist convection) all hint that the strength of vertical mixing changes significantly with latitude, primarily on the length scales of the belts and zones (14). But whereas lightning may be a feature only of the weather layer and the water clouds, disequilibrium species are being dredged from their quench levels, at great depths where temperatures exceed 1000 K. Thus, results from Juno suggest that the cloud-top bands are merely the tip of the iceberg, with atmospheric circulation (horizontal winds and vertical motions) penetrating to great depths.

The unexpected behavior of the volatiles has made constraining Jupiter's deep water abundance even more challenging. Because the signature of water is much smaller than that of ammonia, uncertainties caused by the spatial variability in ammonia severely affect our ability to constrain water. By focusing on the MWR data at the equator where ammonia is well-mixed, the deep water abundance could be constrained to between 1.0 and 5.1 times the solar (O/H) value (17). The uncertainty remains large however, and other regions of the planet have not been explored. We should expect that the distribution of condensing species will be highly variable, with depth and latitude, on the other giant planets as well.

Jupiter and Saturn have one more trick to shield their secrets from view: they are variable with time. Jupiter's bands can expand, contract, fade away and re-appear with spectacular storms, over well-defined multi-year time periods that we are yet to fully understand (18). Saturn exhibits enormous storm outbursts that contribute a significant fraction of the planet's energy budget (19). Do these episodes reflect changes in the deep interior, or are they a consequence of shallower weather-layer processes? As Juno's mission continues, and as Europe prepares the Jupiter Icy Moons Explorer (JUICE) for arrival in 2029, we may finally answer these questions and understand the complex interplay between interior and atmosphere.

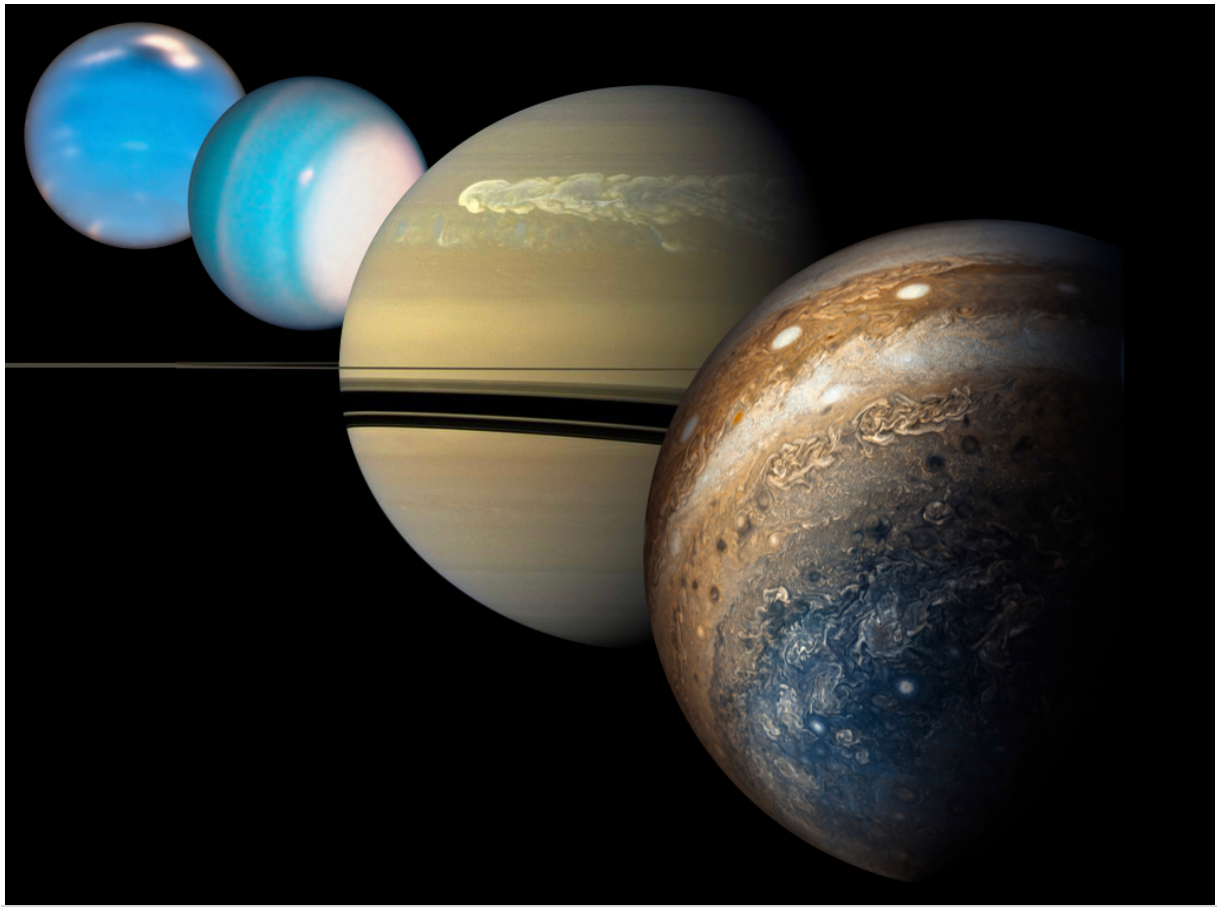

*Figure 1 : The four giant planets shown to scale. From closest to furthest (or right to left) respectively: Jupiter (Juno perijove 6, Credit:NASA/SwRI/MSSS/Gerald Eichstädt/Seán Doran); Saturn during the great storm of 2010-2011 (Cassini, Credit: NASA/JPL-Caltech/Space Science Institute); Uranus and Neptune (Hubble, Credits: NASA/ESA/A. Simon (NASA Goddard Space Flight Center), and M.H. Wong and A. Hsu (University of California, Berkeley)).*

## Uranus and Neptune hold the keys

The last few decades have thus provided tremendous leaps in our understanding of the Gas Giants, whilst the Ice Giants Uranus and Neptune remain poorly explored and mysterious, out in the distant solar system. Ice Giant volatiles show strong equator-to-pole gradients, being massively depleted over the poles and enriched at the equator (20). Clouds are organized into banded patterns, but these are not apparently reflected in wind and temperature contrasts (14). Storms erupt and drift in latitude, and episodic outbursts may teach us about convection in environments where strong density gradients serve to stabilize atmospheric layers, maybe even separating them from the deeper interiors.

Uranus and Neptune's abundant methane clouds and storms have properties similar to those of water clouds in Jupiter and Saturn in terms of abundance and heat content. However, the methane clouds are located at much lower optical depths and are thus much easier to access and study. A mission to Uranus or Neptune including an orbiter with deep remote sensing capability and a probe would enable mapping the deep atmospheric temperature and composition while having a fixed, reliable reference profile. It would thus fully characterize the deep atmosphere of an Ice Giant and constrain its interior structure. This would be key to understand the mechanisms that govern the physics of clouds and storms in planets with hydrogen atmospheres. Indeed, the expanding census of planets beyond our Solar System suggest that Ice-Giant-sized worlds are a common endpoint of the planet formation process, such that future exploration of the Ice Giant atmospheres and interiors, and how they differ from Jupiter and Saturn, is the vital next step in our exploration of the Solar System, filling in the missing link between Gas Giants and terrestrial worlds.

We have begun lifting the veil on the interiors and atmospheres of Jupiter, Saturn. Doing so on Uranus and Neptune is within reach with an ambitious robotic mission to the Ice Giants. It is needed to understand the origin of the Solar System and to analyze with confidence data obtained for the numerous planets with hydrogen atmospheres in our Galaxy.

**Acknowledgements:**

T. Guillot was supported by the *Centre National d'Études Spatiales* and a fellowship from the Japan Society for the Promotion of Science. L.N. Fletcher was supported by a Royal Society Research Fellowship and European Research Council Consolidator Grant (under the European Union's Horizon 2020 research and innovation programme, grant agreement No 723890) at the University of Leicester.

**Author Contributions:**

TG wrote the parts of this article focussing on planetary interiors; LNF wrote parts of this article discussing planetary atmospheres. Both authors reviewed and edited this article.

**Competing Interests:**

The authors declare no competing interests in the preparation of this article.